\documentstyle[multicol,pre,aps,epsfig,amsmath]{revtex}

\newcommand{\A}{A}

\newcommand{\dz}{\partial_z}
\newcommand{\dznull}{\partial_{z_0}}

\begin{document}

\draft

\title{A twist localizes three-dimensional patterns}
\author{A.~G.~Rossberg\footnote{\texttt{http://www.rossberg.net/ag}}}
\address{Department of Physics, Kyoto University, 606-8502 Kyoto, Japan} 
\date{submitted to PRE, April 9, 2000; resubmitted June 26, 2000;
  accepted July 11, 2000} 
\maketitle
\begin{abstract}
  A mechanism for the localization of spatially periodic,
  self-organized patterns in anisotropic media which requires systems
  extended in all three spatial dimensions is presented: When the
  anisotropy axis is twisted the pattern becomes localized in planes
  parallel to the anisotropy axis.  An analytic description of the
  effect is developed and used to interpret recent experiments in the
  high-frequency regime of electroconvection by Bohatsch and
  Stannarius [Phys. Rev. E {\bf 60}, 5591 (1999)].  The localization
  width is found to be of the order of magnitude of the geometrical
  average of pattern wavelength and the inverse twist.
\end{abstract}

\begin{multicols}{2} 
\narrowtext

\section{Introduction}
\label{sec:introduction}

Three-dimensional pattern formation in dissipative systems has been
theoretically investigated to some extent for the case of the complex
Ginzburg-Landau equation
\cite{keener,friri,gaottgu,tornschro,arbikra,roucharay} --- but there
seems to be no experimental model for which the theory is
quantitatively valid.  Qualitatively, many results should also apply
to a similar case, pattern formation in 3D excitable media, for which
controlled experimental realizations exist
(e.g.~\cite{welsh83:_three_belous_zabot,jahnke88:_chemic,vinson97:_contr}),
although the containers are typically rather small.  In both systems
it seems reasonable to concentrate on the structure and dynamics of
the filaments formed by the cores of scroll waves, which largely
control the dynamics in the remaining space.  In a recent theoretical
work, for example, the filament dynamics was investigated for an
anisotropic excitable medium with a twisted anisotropy axis
\cite{wellner00:_predic}, a situation found in the human heart muscle.
It was predicted that the filament typically drifts such as to align
with the anisotropy axis.

Both the complex Ginzburg-Landau equation and excitable media exhibit
traveling waves.  For systems forming \emph{stationarily bifurcating}
spatially periodic structures the dominance of the filaments formed by
topological defects is not so strong, and different effects can be
expected.  A prototype for such a system might be the high-frequency
regime of electroconvection (EC) in nematic liquid crystals.
Convection patterns with a wavelength $\lambda$ which is small
compared to the spatial extensions of the sample are easily obtained
and sustained over long times by applying a sufficiently high ac
voltage (with a modulation frequency higher than the ``cutoff frequency''
$\approx$ the inverse charge relaxation time) at plane, parallel (and
transparent) electrodes enclosing the liquid crystal.  However, there
is still some discussion on whether the patterns are really three
dimensional, i.e.\ {}arising from a bulk instability and forming a
laminar structure, or rather parallel rolls located at the electrodes.

Closely related to this geometrical question is the question about the
mechanism of convection.  The most popular explanation, here called
the standard model (SM), is by the dielectric regime of EC
\textit{via} the Carr-Helfrich mechanism (for details, see the references
below), which is based the Leslie-Ericksen formulation of
nematodynamics.  The SM predicts bulk convection.  It has, on the
linear level, been worked out by Dubois-Violette \textit{et
  al.}~\cite{dvgpp}.  Later it was modified for square-wave driving
voltages \cite{smigaladvdu} and to include flexoelectric effects
\cite{mara,tzk,kbpt} and the effects of finite cell thickness
\cite{ehcbook}.  Recently it was extended to the nonlinear level
\cite{softmodes}.  But attempts to bring the predictions of the SM
(threshold voltage and critical wave number) into quantitative
agreement with experiments were only partly successful.
  
The main challenge to the SM is the ``isotropic'' mechanism for EC
\cite{bablpitr,chipi}, for which the anisotropy of the nematic is only
of secondary relevance.  There, EC is driven by inverted charge
gradients in thin Debye layers near the electrodes, which are not
included in the SM.  The ``isotropic'' mechanism naturally leads to
boundary convection.  Since several material parameters relevant to
the ``isotropic'' mechanism are unknown, only semiquantitative
predictions of the effect can be made.

In principle, both mechanisms are active, but it is reasonable to
assume that one of them dominates.  In a recent effort to settle the
controversy on bulk \textit{vs.}\ {}boundary instability, Bohatsch and
Stannarius \cite{bosta} developed an experiment with a modified
design, which was intended to map the structure across the cell onto
the 2D shadowgraph image (or the diffraction pattern) of light passing
through the cell.


The technique of Bohatsch and Stannarius was to used samples with a
\emph{twisted} geometry: just as in TN LCD displays, the anchoring of
the nematic director $\hat n$ ($\hat n^2=1$) at the upper and lower
electrodes (separation $d$) is such that $\hat n$ is parallel to the
layer but the alignment axis of the director at the upper boundary is
perpendicular to that at the lower boundary.  In the equilibrium
configuration the director is planarly oriented and interpolates
between the boundaries with a uniform twist $\hat n \cdot
\mathop{\mathrm{rot}} \hat n$.  Configurations with positive or
negative twist are thinkable.  They are related to each other by a
reflection.  Without loss of generality a negative twist ($-\pi/2d$)
shall be assumed.  Although inversion symmetry is broken in the
twisted geometry, there is still the symmetry with respect to a
$180^\circ$ rotation of the cell around the axis of the director at
mid plane.  The critical mode at the onset of EC, which breaks the
approximate translation symmetry, can be symmetric with respect to
this rotation (perhaps followed by a translation by half a
wavelength), leading to \emph{normal rolls}, or break also this
discrete symmetry, then leading to \emph{oblique rolls}.
  
In experiments with untwisted cells, the rolls (or laminae) observed
at the threshold of structure formation in the high-frequency regime
are always normal\footnote{When chevron patterns are observed, one is
  not at the threshold of structure formation.} \footnote{In the
  low-frequency regime of EC they can be normal or oblique, however.}.
Thus Bohatsch and Stannarius conclude by an intuitive argument that,
in the twisted cells, the observation of normal rolls at threshold
demonstrates a bulk instability, while the observation of oblique
rolls is an indicator for a boundary instability, in particular when
the rolls are approximately normal to the alignment of $\hat n$ at the
boundaries.  They observe a transition from normal to oblique rolls as
the ac frequency increases.

Assuming that the SM is basically correct,
Sections~\ref{sec:bulk_modes} and \ref{sec:boundary_modes} of this
paper develop the linear theory for the convection pattern in twisted
cells, predicting critical modes localized in planes far from or near
to the electrodes, respectively, depending on the boundary conditions.
Section~\ref{sec:nonlinear_selection} sketches the nonlinear effects
relevant at the threshold of convection and
Section~\ref{sec:conclusion} concludes, in particular with the
interpretation of the experiments in the light of the theory.

\section{Localized bulk modes in the standard model}
\label{sec:bulk_modes}

Below the linear theory of high-frequency EC in twisted cells is
developed within the framework of the SM.  The association of the
obliqueness angle with the position of the convective mode turns out
to be correct.  But it is also found that the experimental results for
twisted cells cannot be transfered to the untwisted cell as it was
suggested by Bohatsch and Stannarius, because the localization of the
linear mode, near the boundaries or in the bulk, is enforced by the
twist itself.

The calculations are based on the general weakly nonlinear
theory~\cite{softmodes} of dielectric EC in large containers, which
has to take the interaction of the convection pattern with several
homogeneous soft modes (hydrodynamic modes) explicitly into account.
One of these modes is the director orientation.  As follows from the
general calculation, a twist does, at lowest order, not excite the
other soft modes of the liquid crystal, such as electric potentials or
fluid motion, even when the electric ac voltage is applied.  The
direct influence of the twist on the convection pattern reduces to a
purely geometric effect (no additional coupling constants not relevant
for the untwisted cell are involved).  The local amplitude and phase
of the convection pattern is expressed by a complex field
$A=A(x,y,z,t)$ such that the modulation of actual physical quantities
$u$ is given by $u(x,y,z,t)=u_0 A(x,y,z,t) \exp(i q_c
x)+c.c.+\text{(higher order corrections)}$, where $q_c$ is the wave
number of the convection pattern at threshold in the untwisted cell
and $u_0$ is a complex constant.  The linear dynamics of $A$ in the
presence of a pure, not too strong twist of the nematic director is
described \cite{softmodes} by
\begin{align}
  \label{thin3da}
  \tau {{{\partial}_t}}{A}=
  \left[ \varepsilon 
    +\xi_x^2 {\partial}_x^2 
    +\xi_y^2 ({\partial}_y- i q_c \varphi)^2
    +\xi_z^2 {\partial}_z^2    
  \right] A.
\end{align}
Here the $z$ axis is perpendicular to the electrodes.  To be specific
assume that $\hat z$ points ``up'' and $-\hat z$ points ``down''.  For
now, the orientation of $x$ and $y$ axis relative to the director
anchoring shall be left unspecified.  This is possible because of the
rotational symmetry underlying Eq.~(\ref{thin3da}) \cite{rohekrpe}.  The
angle between $\hat n$ and the $x$-$z$ plane is denoted by $\varphi$.
Quantitative validity of Eq.~(\ref{thin3da}) requires that $\varphi$
be small and that modulations of $A$ and $\varphi$ in space and time
are slow.  In particular Eq.~(\ref{thin3da}) is never valid across the
whole cell, since $\varphi$ varies over an angle ${\mathcal{O}}(1)$, a
problem that will be discussed below.  The (positive) coherence
lengths $\xi_x$, $\xi_y$, and $\xi_z$ are typically of the order
${{\mathcal{O}}}(q_c^{-1})$, and $\tau$ is of the order of the charge
relaxation time.  The strength of the external stress is expressed by
the reduced control parameter $\varepsilon:=E^2/E_c^2-1$ in terms of
the strength of the external electric ac field $E=V/d$ and the
threshold field for electroconvection $E_c$ for the thick
($d\to\infty$), \emph{untwisted} cell.  $|\varepsilon|$ is assumed to
be small.  (Observe that this definition of $\varepsilon$ differs from
the one used in Ref.~\cite{bosta}, which is based on the threshold in
the \emph{twisted} cell with finite $d$, a quantity to be calculated
below.)  Analytic approximations for all quantities involved in the
description~(\ref{thin3da}) can be found in Ref.~\cite{softmodes}.

The boundary conditions corresponding to Eq.~(\ref{thin3da}) are
\begin{align}
  \label{simple_bc}
  A(x,y,z,y,t)=0\quad\text{at the horizontal boundaries}.
\end{align}

From the translation invariance in the $x$-$y$ plane it
follows that the critical mode is of the form $A=a(z) \exp(i Q x+i P
y)$ with real $Q$ and $P$, i.e.\ {}with a pattern wave vector $\vec
q=(q_c+Q,P)$.  We will \emph{choose} the orientation of the $x$ axis
to be parallel to $\vec q$, in other words $P=0$.  It is easily seen
by separation of variables that according to Eq.~(\ref{thin3da}) the
critical mode, i.e.~the first mode to become unstable as $\varepsilon$
increases, has $Q=0$.  It follows that the twist has no influence on
the wavelength of the critical mode.
  
This result is in agreement with the measurements of Bohatsch and
Stannarius for the cells of intermediate thickness.  But in their
thinnest and in their thickest cell they find shorter wavelengths.
While the deviation in the thin cell can be understood as a breakdown
of the 3D amplitude formalism due to large spatial gradients of $A$
and $\varphi$, the reason for the deviation in the thick cell is not
clear.  In agreement with theoretical expectations the wavelength
decreases with increasing frequency.  The observed deviation from the
approximate $\lambda\sim\omega^{-1/2}$ law at the highest frequencies
indicates that the wavelength has reached the Debye length and charge
diffusion effects need to be taken into account in the theoretical
description, an effect noticed already by Dubois-Violette \textit{et al.}
\cite{dvgpp} (and typically to be expected at the $\sim 10^2$ fold cutoff
frequency \cite{softmodes}).

Choose the origin of the $z$ axis such that $\varphi=0$ at $z=0$.
Denote by $z_0$ the distance of the lower boundary from $z=0$ (i.e.\ 
{}the lower boundary is at $z=-z_0$).  By the preceding construction,
the amplitude of the critical mode is a function of $z$ alone and
satisfies
\begin{align}
  \label{schroedinger}
  0= \left[ \varepsilon -\xi_y^2 q_c^2 {\varphi}^2 +\xi_z^2
    {\partial}_z^2 \right] \A
\end{align}
with $\varphi=\pi z/2 d$. The equation has the form of the
time-independent Schr{\"o}\-din\-ger equation for the quantum harmonic
oscillator.

Ignore, for a moment, the horizontal boundaries, and impose instead
the B.C.\ 
\begin{align}
  \label{bounded_bc}
  A\quad\text{is bounded as}\quad |z|\to\infty.
\end{align}
Then the lowest eigenvalue of Eq.~(\ref{schroedinger}) is
\begin{align}
  \label{threshold}
  \varepsilon=\varepsilon_c:=\frac{\pi\, q_c\, \xi_y\, \xi_z}{2\, d},
\end{align}
with a critical mode 
\begin{align}
  \label{critical_mode}
  \A=\exp\left( -\frac{z^2}{2\, w^2}\right)
\end{align}
with a width
\begin{align}
  \label{width}
  w=\left(\frac{2\, d\, \xi_z}{\pi\, q_c\, \xi_y}\right)^{1/2}.
\end{align}
Equations~(\ref{threshold},\ref{critical_mode}) correspond to the
ground state of the harmonic oscillator.

Now assume $d \, q_c \gg 1$, which is required for the validity of
Eq.~(\ref{thin3da}).  Then $\varepsilon_c={\mathcal{O}}(d \,
q_c)^{-1}\ll 1$ and, within the range $|z|={\mathcal{O}}(w)$ covered
by the critical mode, $\varphi={\mathcal{O}}(w/d)={\mathcal{O}}(d \,
q_c)^{-1/2}\ll 1$.  Thus, for $|z|={\mathcal{O}}(w)$, the
solution~(\ref{threshold},\ref{critical_mode}) is a physically valid
solution of Eq.~(\ref{thin3da}).  For $|z|\gg w$ the critical mode is
to high accuracy zero and thus a solution of the linear problem, also
beyond the range of validity of Eq.~(\ref{thin3da}).  Of course any
homogeneous B.C., in particular B.C.~(\ref{simple_bc}), are satisfied,
provided the boundaries are sufficiently far away.
Equations~(\ref{threshold},\ref{critical_mode}) therefore give a
physically valid threshold and critical mode.

Next, consider the situation that one of two boundaries, e.g.~the
lower one, is near $z=0$ [i.e.~$z_0={\mathcal{O}}(w)$], the other
boundary far away.  Then the lower B.C.\ {}becomes
\begin{align}
  \label{fixed_bc}
  \A=0\quad\text{at}\quad z=-z_0
\end{align}
while the upper B.C.\ {}remains effectively
\begin{align}
  \label{open_bc}
  A\quad\text{is bounded as}\quad z\to\infty.
\end{align}
This eigenvalue problem has to be solved numerically (after rescaling,
it depends only on the parameters $\varepsilon/\varepsilon_c$ and
$z_0/w$); here a shooting method is used.  The lowest eigenvalue
$\varepsilon(z_0)$ as a function of $z_0$ is shown in
Fig.~\ref{fig:epsilon}b (lower curve).  As $z_0$ decreases to zero and
the maximum of the critical mode approaches the boundary, the Gaussian
shape of $A(z)$ is deformed and the threshold $\varepsilon(z_0)$
increases monotonically.  In particular one finds $\varepsilon(0)=3\,
\varepsilon_c$, the eigenvalue of the first excitation and lowest
antisymmetric mode of the ``harmonic oscillator'', with the
corresponding eigenmode.  Excitations near the boundaries have higher
thresholds than in the bulk and are inhibited.

Thus the theory predicts that in thick enough cells the critical
mode is localized within a horizontal layer which
covers only a small fraction $={\mathcal{O}}(w/d)={\mathcal{O}}(q_c
d)^{-1/2}$ of the cell.  The orientation of the critical wave vector
$\vec q$ and the position of the maximum of the critical mode within
the sample are coupled, such that $\vec q$ is parallel to $\hat n$ at
the maximum of $A(z)$.  The orientation of the critical wave vector
$\vec q$ lies between the two orientations of director anchoring at
the boundaries.  Small angles of order ${\mathcal{O}}(w/d)$ between
$\vec q$ and director anchoring are forbidden, but otherwise the
orientation of $\vec q$ with respect to the symmetry axis, i.e.\ {}the
obliqueness angle, is undetermined.  One is thus dealing with a
\emph{highly degenerate problem} with critical wave vectors lying on a
circular arc, similar to pattern formation in isotropic systems, where
the critical wave vectors lie on a circle.
  
In thin cells, where $w/d,\,d\,q_c={\mathcal{O}}(1)$, the
three-dimensional amplitude equation~(\ref{thin3da}) is not an
accurate description anymore.  But when solved numerically with
B.C.~(\ref{simple_bc}) as a qualitative model, it predicts normal
rolls in agreement with experiments.

The theoretical analysis for thick cells seems to correspond well with
the observations of Bohatsch and Stannarius in the ``broad''
intermediate frequency range (within the high-frequency regime),
where, near onset, the distribution of the wave-vector orientations is
``broad, smeared out, and fluctuating'' \cite{bosta}.  Unfortunately,
no experimental raw data for this range have been published, yet,
which makes it hard to judge how well defined the maxima of these
distributions are, which they \emph{do} nevertheless find at small but
finite obliqueness angles ($\approx 10^\circ$), and which are
difficult to understand from the theory presented here.

\section{Boundary modes with modified B.C.}
\label{sec:boundary_modes}

The experimental observation that, for frequencies above the
intermediate ``transition range'', the distribution of the wave
vectors near onset becomes narrower with maxima at obliqueness angles
near $\pm45^\circ$, i.e.\ {}parallel to the surface anchoring, seems to be
in contradiction with the theory based on the SM.
  
A simple explanation, which should be examined experimentally before
coming to a final conclusion, would be the combination of two effects.
The first effect is that, at higher frequencies, where the pattern
wavelengths are shorter and wave-optical effects have to be taken into
account, the optical contrast of the shadowgraph method and the
intensity of the reflexes in the optical far field become weaker.  As
a result, the apparent optical threshold of convection could lie
slightly above the actual threshold.  Wave-optical effects become
important at pattern wavelength $\lambda$ of the order of magnitude of
the geometrical average of the optical wavelength and the thickness of
the convective layer $={\mathcal{O}}(w)={\mathcal{O}}(\lambda
d)^{1/2}$ \cite{waveoptics}, a condition satisfied in the present
case.  The second effect is that the value of the reduced control
parameter corresponding to the onset of nontrivial, nonlinear behavior
involving homogeneous soft modes scales like $(\lambda/d)^2$ or even
$(\lambda/d)^4$ \cite{softmodes}, and thus decreases with increasing
frequency.  In combination of these two trends, the observed large
obliqueness angles near threshold in thick cells at high frequencies
might, in fact, correspond to already fully developed nonlinear
effects, such as the chevron pattern.  In particular the obliqueness
angles above $45^\circ$ (Ref~\cite{bosta}, Fig.~11,
$3500\,\mathrm{Hz}$) would then be easier to understand.

But one could also try to explain the observation of preferred
obliqueness angles near $\pm45^\circ$ as a result of some mechanism of
wave-vector selection that removes the degeneracy of the problem.
Mostly linear selection mechanisms {\it via\/} perturbations from the
boundaries shall be considered here.  Nonlinear mechanisms will
shortly be discussed in Section~\ref{sec:nonlinear_selection}.

If there is a linear selection mechanism for large obliqueness angles
at threshold, at least one of the tree
conditions~(\ref{schroedinger},\ref{fixed_bc},\ref{open_bc}) must be
invalid.  The reason may be found in the truncation errors of the
asymptotic expansion which leads to
Eqs.~(\ref{schroedinger},\ref{fixed_bc},\ref{open_bc}), the excluded
effect of noise, or an insufficient description on the hydrodynamic
level.  All possible truncation errors become, by one way or another,
large only for $z_0={\mathcal{O}}(d)$.  This will be demonstrated for
one case below.  If noise is the main cause for selecting $z_0$ (which
is presumably the case at intermediate frequencies), it is reasonable
to assume $z_0$ to be randomly distributed over $[0,d]$, leading again
to $z_0={\mathcal{O}}(d)$ on the average.  Thus it can be concluded
that, in fact, the description on the hydrodynamic level, in
particular near the boundaries, is insufficient.

For the eigenvalue problem considered above, the B.C.~(\ref{fixed_bc})
turns out to be a somewhat singular case of the more general condition
\begin{align}
  \label{mixed_bc}
  \A=\mu \dz \A\quad\text{at}\quad z=-z_0
\end{align}
with a real parameter $\mu$. To discuss the implications of $\mu\ne0$ as
one possible truncation error or as a possible outcome of an extended
hydrodynamic description, the eigenvalue problem~(\ref{schroedinger})
with B.C.~(\ref{open_bc},\ref{mixed_bc}) shall now
be investigated.  This will also yield an analytic argument, robust
with respect to perturbations, that supports the numerical observation
that for $\mu=0$ the function $\varepsilon(z_0)$ is monotonously
decaying with $z_0$.

Up to normalization, there is, for fixed $\varepsilon$, a single
nontrivial solution $\A(z)=f(\varepsilon,z)$ of
Eq.~(\ref{schroedinger}) satisfying the upper B.C.~(\ref{open_bc}).
The lower B.C.~(\ref{mixed_bc}) determines a discrete set of $z_0$ for
given $\varepsilon$ or, respectively, $\varepsilon=\varepsilon(z_0)$
is determined implicitly by the lower B.C.~as a multi-valued function
of $z_0$, i.e., by
\begin{align}
  \label{mixed_eps_bc}
  f(\varepsilon(z_0),-z_0)=\mu f^\prime(\varepsilon(z_0),-z_0),
\end{align}
where the prime on $f$ denotes a differentiation with respect to the
second argument.

At a local minimum of $\varepsilon(z_0)$ one has
$\dznull\varepsilon(z_0)=0$, and differentiation of
Eq.~(\ref{mixed_eps_bc}) with respect to $z_0$ yields
\begin{align}
  \label{diff_bc}
  f^\prime(\varepsilon(z_0),-z_0)=\mu
  f^{\prime\prime}(\varepsilon(z_0),-z_0).
\end{align}
By combining Eqs.~(\ref{mixed_eps_bc},\ref{diff_bc}) and the defining
equation of $f$, Eq.~(\ref{schroedinger}), one arrives at
\begin{align}
  \label{algebraic1}
  \xi_z^2 \, f=\mu^2 \left[\frac{\pi^2 \xi_y^2 q_c^2 z_0^2}{4 d^2}
    -\varepsilon\right]\,f.
\end{align}
For nonzero $\mu$ the common factor $f$ can be canceled since $f\ne0$
by Eq.~(\ref{mixed_eps_bc}).  A simple algebraic relation between
$\varepsilon$ and $z_0$ (which determines the wave vector) for the
critical mode is thus obtained.

As $\mu \to 0$, two cases can be distinguished.  In the first case
$z_0$ remains bounded.  Then Eq.~(\ref{algebraic1}) reduces
asymptotically to
\begin{align}
  \label{boundary_mode}
  \varepsilon \sim -\xi_z^2/\mu^2.
\end{align}
Corresponding solutions exists only for $\mu<0$. They can be
approximated by $f(\varepsilon,z)\approx \exp(z/\mu)$ and are strongly
localized near the lower boundary.  But due to the large values of
$|\varepsilon|$ (as $\mu\to0$) these solutions can not be expected to
be always physically meaningful.  In the second case $z_0$ increases over all
bounds.  Then small deviations from eigenvalue~(\ref{threshold}) and
solution~(\ref{critical_mode}) are sufficient to satisfy the B.C..
Thus $\varepsilon$ remains bounded and Eq.~(\ref{algebraic1}) reduces
asymptotically to
\begin{align}
  \label{asymptotic}
  z_0\sim\frac{2 d \xi_z}{\pi q_c \xi_y |\mu|}=\frac{w^2}{|\mu|}.
\end{align}
In particular $z_0\to\infty$ as $\mu \to 0$, supporting the numerical
observation that $\varepsilon(z_0)$ is strictly monotonously decaying
for $\mu=0$.
  
It is easily verified that the analytic argument for $z_0\to\infty$ as
$\mu \to 0$ is, as far as physically interesting, robust against
additions of higher order corrections of the form $g(\varphi^2) A$ to
the r.h.s.\ {}of Eq.~(\ref{schroedinger}) or the replacement of the
term $\xi_z^2 \dz^2 A$ by some more precise form $h(\varphi^2) \dz^2
A$ [assuming that $h(\varphi^2)$ has no zeros for real $\varphi$,
i.e., the PDE be nonsingular].  All other higher order corrections to
Eq.~(\ref{schroedinger}) involve higher order derivatives (and
additional boundary conditions).  If these corrections are small they
are, in a first approximation, active only in a thin boundary layer
and lead, effectively, to modified boundary conditions of the
form~(\ref{mixed_bc}).  Thus, taking the validity of the upper
B.C.~(\ref{open_bc}) for granted, a modified lower B.C.\ {}is always
involved in a localization near the lower boundary.

\begin{figure}[p]
  \begin{center}
    \epsfig{file=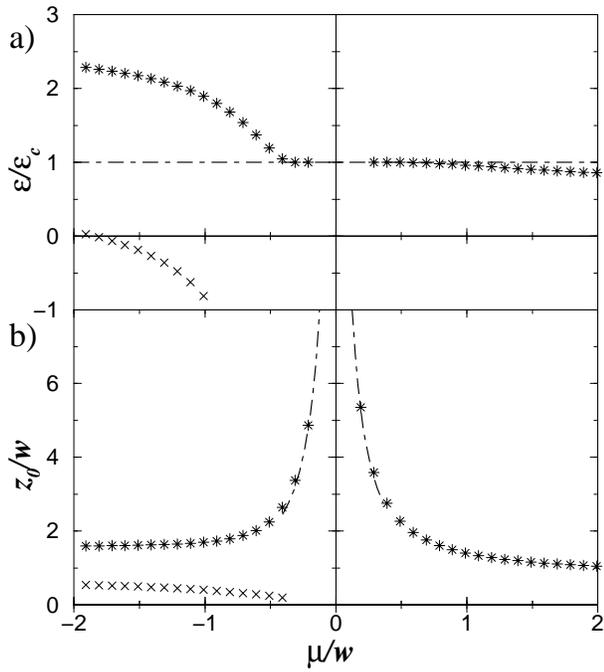,width=8 cm}
    \caption{(a) The threshold value of dielectric electroconvection 
      (relative to the bulk threshold in an untwisted nematic) in
      units of $\varepsilon_c$ [Eq.~(\ref{threshold})] and (b) $z_0$
      ($\approx$ distance of the maximum of the critical mode from the
      boundary) in units of the width $w$ of the critical mode
      [Eq.~(\ref{width})].  Both as functions of the parameter $\mu$
      in the boundary condition~(\ref{mixed_bc}).  Stars and crosses
      are numerical solutions.  The crosses correspond to a thin
      boundary mode.  The dashed lines are the analytic approximations
      for small $|\mu|$, Eq.~(\ref{asymptotic}) for (b).}
    \label{fig:numerics}
  \end{center}
\end{figure}

\begin{figure}[p]
  \begin{center}
    \epsfig{file=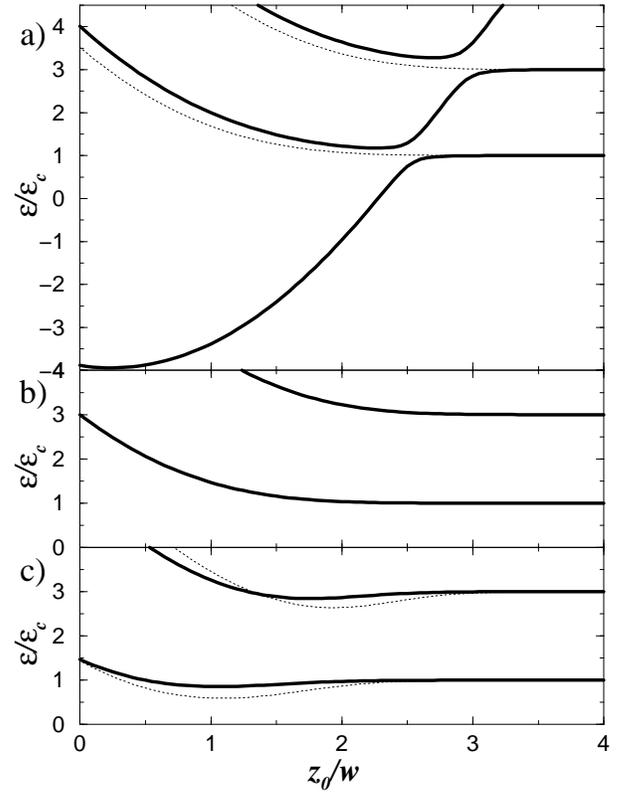,width=8 cm}
    \caption{The lowest eigenvalues $\varepsilon(z_0)$ for (a)
      $\mu=-0.5\,w$,
      (b) $\mu=0$, and (c) $\mu=2.0\,w$.  The solid lines correspond to
      the eigenvalue problem
      Eqs.~(\ref{schroedinger},\ref{fixed_bc},\ref{open_bc}), the
      dotted lines to a modified model that does not contain the
      boundary mode leading to the large negative eigenvalues
      $\varepsilon(z_0)$ for negative $\mu$. For $\mu=0$ solid and
      dotted lines are identical.}
    \label{fig:epsilon}
  \end{center}
\end{figure}

To better understand the situation for $\mu\ne0$, a numerical solution of
the problem is required.  Figure~\ref{fig:numerics} shows numerical
results for the values of $\varepsilon$ and $z_0$ at threshold [i.e.\ 
{}the minimum of $\varepsilon(z_0)$] for finite $\mu$.  Although in
the limit $\mu\to0$ numerics break down, the asymptomatic results are
reproduced for small $\mu$.  The calculations also show that for
$\mu>0$ the minima of $\varepsilon(z_0)$ lie below the bulk ($z_0\to
\infty$) value $\varepsilon_c$ as expected, but for $\mu<0$ are, on
the starred branch (s.\ {}Fig.~\ref{fig:numerics}), above
$\varepsilon_c$.  A look at the function $\varepsilon(z_0)$
(Fig.~\ref{fig:epsilon}) shows that for negative $\mu$ the minima on
the starred branch belong to the branch of $\varepsilon(z_0)$ which
approaches $3 \varepsilon_c$ as $z_0\to \infty$.  It becomes also
clear that the existence of the minima of $\varepsilon(z_0)$ for $\mu<0$
is closely related to the existence of the thin boundary mode
corresponding to the eigenvalue~(\ref{boundary_mode}): The minima
result from the avoided crossings of the spectrum of the modes
localized near $z=0$, which have decreasing eigenvalues for increasing
$z_0$, and the eigenvalue of the thin boundary mode, which increases
with $z_0$.  If the thin boundary mode is an unphysical artifact of
the boundary conditions the minima in the function $\varepsilon(z_0)$
are also unphysical.  This point can be demonstrated by using the
original B.C.~(\ref{fixed_bc}) as the lower boundary condition and
adding, to model the mixed B.C.~(\ref{mixed_bc}), a term $b(z+z_0) \dz
\A$ to the l.h.s.~of Eq.~(\ref{schroedinger}), where $b(z)$ is, e.g.,
given by
\begin{align}
  \label{b_of_z}
  b(z)=
  \left\{
    \begin{array}[c]{ll}
      13.2\, \varepsilon_c\, w\, \mu & \text{if}\quad z< 2.83\, w\\
      0 &\text{otherwise}.
    \end{array}
  \right.
\end{align}
With these modifications the thin boundary mode --- and therefore the
avoided crossings and the minima of $\varepsilon(z_0)$ --- are suppressed
for $\mu<0$, but otherwise the spectra change only little (s.
Fig.~\ref{fig:epsilon}).

Can higher order corrections to the boundary conditions that arise
from the asymptotic expansion \emph{in the framework of the SM}
explain the preference of large obliqueness angles?  From the figure
in Ref.~\cite{softmodes} displaying the hydrodynamic boundary layer of
dielectric electroconvection, a value $\mu\approx -0.4 \,q_c^{-1}$ can
be obtained.  The sign follows directly from the fact that convection
is suppressed by the no-slip boundary conditions for the hydrodynamic
velocity field, the order of magnitude follows from the geometry of
the flow.  By the sign of $\mu$ a preference of large obliqueness
angles (boundary instability) is excluded.  Even if the sign would
change but the magnitude remain the same, Eq.~(\ref{asymptotic}) would
give an unphysical value $z_0>d$.

\section{Nonlinear selection}
\label{sec:nonlinear_selection}

In principle a nonlinear selection of large obliqueness angles is also
thinkable.  At threshold, two kinds of nonlinear mechanisms are active
\cite{softmodes}.  The first is the direct, local, nonlinear
saturation of the pattern amplitude, the strength of which is measured
by the coefficient $g$ in Ref.~\cite{softmodes}.  This mechanism
inhibits the coexistence of linear modes of the
form~(\ref{critical_mode}) if the overlap of the two modes along $z$
is too large.  Correspondingly, the wave vectors of coexisting linear
modes can be expected to enclose some finite minimum angle.  This
could, for not too small $w/d$, lead to the preference of certain
wave-vector combinations which fill the allowed range of $\approx
\pm45^\circ$ particularly well, but it does not imply that large
obliqueness angles are generally preferred.
  
The second nonlinear mechanism acts indirectly \textit{via} the
electric potential.  At positions in the sample where $|A|^2$ is
large, additional charge transport processes are active, which
effectively increase ($S_E$ in Ref.~\cite{softmodes} is positive) the
electrical conductivity.  Consequently, the strength of the electric
driving field is reduced where $|A|^2$ is large and increased where
$|A|^2$ is small, under the constraint that the total voltage drop is
constant.  But this effect is typically weak compared to direct
nonlinear suppression, and presumably does not change the qualitative
properties of the planform selection problem.

\section{Conclusion}
\label{sec:conclusion}

It was shown that a twist of the nematic director -- or another
anisotropy axis in a different system -- leads to a localization of
the pattern in planes parallel to the director.  When the twist is not
too strong, the localized linear modes can be calculated in the
framework of the amplitude formalism.  In electroconvection, the $z$
axis in Eq.~(\ref{thin3da}) plays, apart form being the twist axis,
another special role -- the external electric field is parallel to
$\hat z$.  But little would change if this additional anisotropy was
missing (e.g.\ {}for reaction-diffusion patterns in uniaxially
anisotropic media).  Only $\xi_y$ and $\xi_z$ would then be equal by
symmetry.

For dielectric EC in twisted nematic (TN) cells the localization width
$w={\mathcal{O}}(d/q_c)^{1/2}$ is given by Eq.~(\ref{width}).  When the
material parameters of the liquid crystal are known, $w$ can be
calculated using the analytic approximations for the coherence lengths
and the wave number from Ref.~\cite{softmodes}.
As a result of the localization, the critical wave vector is
degenerate on a circular arc.  The angle enclosed by the arc might be
increased beyond $90^\circ$ by using supertwist nematic cells (STN).
One does then obtain a pattern-forming system with a geometry very
similar to a two-dimensional isotropic system, but with a different
topology (no invariance of the pattern under $180^\circ$ rotation) and
very particular nonlinear interactions.

The observation of normal rolls in thin cells and at low frequencies
(large wavelength) by Bohatsch and Stannarius \cite{bosta} is
compatible with --- and presumably predictable from --- the standard
model.  The transition to a state with a ``broad, smeared out, and
fluctuating'' distribution of wave-vector orientations at higher
frequencies can be understood from the wave-vector degeneracy.  The
theory also reproduces their observation that the transition takes
place at some more or less fixed ratio $\lambda/d$.  This corresponds
to some fixed ratio $w/d$ at which the B.C.~(\ref{bounded_bc}) become
appropriate and the degeneracy sets in.  Assuming that the
observations at yet higher frequencies, where the wave vectors were
better defined and aligned with the director anchoring, did not suffer
from the technical difficulties described in the beginning of
Sec.~\ref{sec:boundary_modes}, they seem to exhibit effects not
described by the standard model.  These effects seem to lead
effectively to modified boundary conditions of the
form~(\ref{mixed_bc}) with $\mu>0$.  Nevertheless this does not imply
that, just as in the twisted cell, there are boundary modes also in
the untwisted cell.  Solution of the linear problem for the untwisted
cell with B.C.~(\ref{mixed_bc}), $\mu>0$, and a corresponding upper
B.C.\ {}yields a critical mode covering the whole cell.  Since the
theory can explain the transition at the lower end of the
intermediate-frequency range with fluctuating wave vector but does
not imply that there is another transition at an upper bound, it may
be useful to investigate the two transitions independently also in 
experiments.  The intermediate-frequency range is probably an
independent regime, and not just as the crossover between the
low-frequency and the high-frequency range.

Conclusive information on the 3D structure of the high-frequency mode
in untwisted cells might be obtained by investigating spectrally
resolved shadow graphs or light scattering.

It is my pleasure to thank R.~Stannarius for useful comments on the
manuscript, Y.~Kuramoto and the Kyoto University for providing a
working environment, and the Japan Foundation for the Promotion of
Science (P98285) and the Ministry of Education, Science, Sports and
Culture in Japan for financial support.

\pagebreak{}


\end{multicols}

\end{document}